# On the Impact of Dimension Reduction on Graphical Structures

Fang Han[*], Huitong Qiu[†], Han Liu[‡], and Brian Caffo[§]

October 13, 2014

## Abstract

Statisticians and quantitative neuroscientists have actively promoted the use of independence relations for investigating brain networks, genomic networks, and other measurement technologies. Estimation of these graphs depends on two steps. First is a feature extraction by summarizing measurements within a region of interest to create nodes. Secondly, these summaries are used to create a graph representing relations of interest. In this manuscript we study the impact of dimension reduction on graphs that describe different notions of relations between random variables. We are particularly interested in undirected graphs that capture the random variables' marginal and conditional independence relations. A dimension reduction procedure can be any mapping from a $d$ dimensional Euclidean space to a $K$ dimensional Euclidean space. $K$ is typically much smaller than $d$. We exploit a general framework for modeling the raw data and advocate that in estimating the undirected graphs, acceptable dimension reduction procedures should be graph-homotopic. In other words, the graphical structure of the data after the dimension reduction should inherit the main characteristics of the graphical structure of the raw data. We show that, in terms of inferring undirected graphs that characterize the conditional independence relations between random variables, many dimension reduction procedures, such as the mean, median, or principal components, cannot be theoretically guaranteed to be graph-homotopic. The implications of this work are broad. In the most charitable setting for researchers, where the correct node definition is known, graphical relations can be contaminated merely via the dimension reduction. In the end, confined in the Gaussian graphical model, we give a broad analysis on linear dimension reduction methods. We characterize necessary and sufficient conditions for those dimension reduction procedures to be graph-homotopoic. We also illustrate the impact of dimension reduction on graphical structures by studying a real neuroimaging data.

[*]Department of Biostatistics, Johns Hopkins University, Baltimore, MD 21205, USA; e-mail: fhan@jhu.edu
[†]Department of Biostatistics, Johns Hopkins University, Baltimore, MD 21205, USA; e-mail: hqiu@jhu.edu
[‡]Department of Operations Research and Financial Enginerring, Princeton University, Princeton, NJ 07833, USA; e-mail: hanliu@princeton.edu
[§]Department of Biostatistics, Johns Hopkins University, Baltimore, MD 21205, USA; e-mail: bcaffo@jhu.edu
[¶]We thank for very helpful discussions with Xi Luo and Moo K. Chung.

# 1 Introduction

Many modern datasets, such as those from genomic microarrays, digital photographs, and functional magnetic resonance imaging scans, are often high dimensional. For the purpose of computing and scientific interpretation, the datasets' dimensionality often needs to be reduced. A dimension reduction procedure, in general, is a mapping from a $d$ dimensional Euclidean space to a $K$ dimensional Euclidean space, with $K$ usually much smaller than $d$. The dimension reduced random vector is expected to have similar topological structure as the raw data. Retaining different topological structure results in different dimension reduction procedure.

In the unsupervised learning literature, Fodor (2002) summarized the dimension reduction procedures, including principal component analysis, factor analysis, projection pursuit, and independent component analysis. We refer to her manuscript for a thorough review. In the supervised learning literature, the idea of sufficient dimension reduction is often exploited. There the target is to find a low dimensional vector, typically a linear combination of the original predictors, such that the relation between the response and the original predictors can be preserved as much as possible (Cook et al., 2007). In the graphical model literature, when some prior information on the statistical or geometric properties of the raw data, summarized as a certain notion of a graph, is available, a general framework for dimension reduction called graph embedding was proposed for best preserving the characteristics of the graph (Yan et al., 2007).

This manuscript studies a general problem about dimension reduction. That is, what is the impact of dimension reduction on the graphs describing the marginal and conditional independence relations between random variables? The implications of the research are broad. For example, in brain imaging, node definition is a complex task having many compromises of practicality. This work investigates the setting where an oracle node definition is known at the onset, yet unacknowledged problems still persist.

Suppose that we have a $d$ dimensional random vector $\widetilde{X} = (X_1, \ldots, X_d)^T \in \mathbb{R}^d$. An undirected graph, $G = (V, E)$, is a graph with the vertex set $V = \{1, \ldots, d\}$ and edge set $E \subset \{(j, k) : j, k = 1, \ldots, d\}$. Graph $G$ describes the marginal or conditional independence relations of the random variables $X_1, \ldots, X_d$. Take the marginal independence relations for example. There we have $(j, k) \notin E$ if and only if $X_j$ is independent of $X_k$.

A dimension reduction procedure is a surjective mapping $g(\cdot) : \mathbb{R}^d \to \mathbb{R}^K$. We define $\widetilde{Y} = (Y_1, \ldots, Y_K)^T = g(\widetilde{X})$ to be the dimension reduced random vector of $\widetilde{X}$. If we study $G$ based on $\widetilde{Y}$, then an acceptable mapping, $g(\cdot)$, should force $\widetilde{Y}$ to inherit as many characteristics encoded in the graphical structure of $\widetilde{X}$ as possible.

In this manuscript we are particularly interested in evaluating the impact of dimension reduction on graphical structures under a specific scheme. Suppose that we have $K$ non-overlapping clusters of nodes, indexed by $C_1, \ldots, C_K \subset \{1, \ldots, d\}$. The dimension reduction procedure summarizes the random variables in each cluster $\widetilde{X}_{C_i}$ using one random variable $Y_i$. In this setting, we define a dimension reduction procedure to be graph-homotopic if the graphical structure of $Y_1, \ldots, Y_K$ is identical to the block graphical structure of $\widetilde{X}_{C_1}, \ldots, \widetilde{X}_{C_K}$.

One motivating application of this study is functional magnetic resonance imaging data analysis for estimating functional connectivity networks in the brain. Such a network characterizes the brain's connectivity, where the vertices in the graph represent regions of interest in the brain, and the edges in the graph represent the binary correlation strengths (Rubinov and Sporns, 2010; Power et al., 2011). Accordingly, although not explicitly stated, all the existing estimation procedures of the brain connectivity network can be summarized in two steps (Power et al., 2012; Lee et al., 2013; Eloyan et al., 2012; Han et al., 2013; Power et al., 2013). First, there is a dimension reduction step. For registered and processed high resolution brain imaging data, we first define the regions of interest, and then summarize voxels in each region to one value via a dimension reduction method. Secondly, there is an inference step. That is, estimate a certain notion of graphs based on the dimension reduced data obtained in the first step.

In this case, the clusters, $C_1, \ldots, C_K$, are the regions of interest. In estimating the graph of the dimension reduced data, one is intrinsically assuming that it can inherit the main characteristics of the graph of the non-reduced data. We emphasize that by assuming the existence of $C_1, \ldots, C_K$, we assume a charitable circumstance for neuroscience researchers in having a correctly known parcellation.

Brain connectivity networks are regularly estimated using the correlation and partial correlation measures. See, for example, Zalesky et al. (2012) and the references therein. When $\widetilde{X}$ is Gaussian, the correlation and partial correlation graphs correspond to the marginal and conditional independence graphs. Marginal independence is more widely used in the area. However, conditional independence is conceptually preferable, or at least an important complementary approach, given its exploration of multivariate relations (Lauritzen, 1996). We notice that there has been a vast literature in promoting the use of conditional independence graphs in neuroscience. We refer to Smith et al. (2011), Varoquaux et al. (2012), Ryali et al. (2012), and the references therein, for the advantages in interpretability.

The major contribution of this manuscript is to propose the concept of graph-homotopic mapping, and show that, with regard to the conditional independence relations, any dimension reduction method cannot be theoretically guaranteed to be graph-homotopic. This alerts researchers to carefully design dimension reduction methods while studying graphical structures. More detailed recommendations are put in Section 4.1.

In the end, we show that there exist graph-homotopic mappings for a subset of graphs. We focus on the general factor model. We characterize necessary and sufficient conditions for the latent factors to inherit the graphical structure of the raw data. In particular, we propose two models of interest, under which the dimension reduced random variables, $Y_1, \ldots, Y_K$, as the latent factors of $\widetilde{X}_{C_1}, \ldots, \widetilde{X}_{C_K}$, have consistent graphical structure to that of $\widetilde{X}$.

## 2 Graph-Homotopic Mapping

### 2.1 Basic Definitions

In this section we study the graph-homotopic mappings associated with the marginal and conditional independence graphs. In the sequel, we adopt the usual notation (Dawid, 1979) of marginal and conditional independence. For any three random vectors $\widetilde{X}, \widetilde{Y}, \widetilde{Z}$, we write

$$\widetilde{X} \perp \widetilde{Y} \text{ if and only if } \widetilde{X} \text{ is independent of } \widetilde{Y},$$
$$\widetilde{X} \perp \widetilde{Y} \mid \widetilde{Z} \text{ if and only if } \widetilde{X} \text{ is independent of } \widetilde{Y} \text{ conditioning on } \widetilde{Z}.$$

Throughout the paper we focus on the following model. Let $\widetilde{X} = (X_1, \ldots, X_d)^T$ be a $d$ dimensional random vector. We define the undirected graph $G_R = (V_R, E_R)$ as follows. The set $V_R$ consists of all vertices $\{1, \ldots, d\}$ corresponding to $X_1, \ldots, X_d$, and $E_R$ represents the set of edges satisfying $(j, k) \in E_R$ if and only if $(k, j) \in E_R$. Here $d$ can be regarded as the number of voxels in the brain images.

In parallel to the definition of region of interest in brain network analysis, we assume there are $K$ clusters in $\{X_1, \ldots, X_d\}$. For each $j \in \{1, \ldots, K\}$, let $C_j$ be the index set of all covariates in the $j$th cluster and $d_j$ be the dimension of $\widetilde{X}_{C_j}$. Here $K$ can be regarded as the number of regions of interest. We assume that there is no overlap among $\{C_1, \ldots, C_K\}$. Further, without loss of generality, we assume the nodes are ordered such that $\widetilde{X}$ can be written as $(\widetilde{X}_{C_1}^T, \ldots, \widetilde{X}_{C_K}^T)^T$. We are interested in the connectivity structure of the clusters $\{\widetilde{X}_{C_1}, \ldots, \widetilde{X}_{C_K}\}$, namely the undirected graph $G_C = (V_C, E_C)$. The set $V_C$ consists of vertices $\{1, \ldots, K\}$ corresponding to $\widetilde{X}_{C_1}, \ldots, \widetilde{X}_{C_K}$, and the set of edges $E_C$ is defined as

$$(j, k) \notin E_C \text{ if and only if } (i_j, i_k) \notin E_R \text{ for all } i_j \in C_j \text{ and } i_k \in C_K.$$

Let $\widetilde{Y} = (Y_1, \ldots, Y_K)^T$ be the dimension reduced random vector of $\widetilde{X}$, with $Y_j$ corresponding to the $j$th cluster, $\widetilde{X}_{C_j}$. Let $G_{DR} = (V_{DR}, E_{DR})$ be the graph of $\widetilde{Y}$, where $V_{DR} = \{1, \ldots, K\}$ represents the set of vertices corresponding to $Y_1, \ldots, Y_K$, and $E_{DR}$ is the set of edges for $\widetilde{Y}$. The notions of dependence relations between random variables described by graphs $G_R, G_C$, and $G_{DR}$ should be the same. We advocate that, if one is interested in estimating the graph, $G_C$, any acceptable dimension reduction procedure should be a graph-homotopic mapping from $\mathbb{R}^d$ to $\mathbb{R}^K$. The definition of graph-homotopic mapping is as follows.

**Definition 2.1.** A dimension reduction procedure is a graph-homotopic mapping if and only if the graph, $G_{DR}$, of the dimension reduced random vector, $\widetilde{Y}$, is identical to the graph $G_C$, in the raw random vector, $\widetilde{X}$, i.e., $E_{DR} = E_C$. And we call the dimension reduction procedure a weakly graph-homotopic mapping if $E_{DR} \subset E_C$.

We emphasize that, intrinsically, the above setting indicates two points. (i) The graph, $G_R$, has many dense subgraphs, called cliques. (ii) The locations of these cliques are known, so that the dimension reduction procedure could be conducted. Such a setting is a reasonable starting point in, for example, brain connectivity analysis. First, the resting state functional magnetic resonance image intensities are highly correlated within certain functional systems, such as the visual or motor system, and less correlated between different functional systems (Biswal et al., 1995; Dosenbach et al., 2007; Nelson et al., 2010). Secondly, the information about the locations of these subgraphs can be obtained via combining biological information regarding brain functional specialization with information learned using state-of-the-art data analysis tools (Margulies et al., 2007; Power et al., 2011).

We consider two types of graphs that are routinely exploited in exploring brain connectivity networks, the marginal independence graph and the conditional independence graph. The formal definition is as follows.

**Definition 2.2.** We call $G = (V, E)$ a marginal independence graph if

$$(j, k) \notin E \quad \text{if and only if} \quad X_j \perp X_k.$$

We call $G$ a conditional independence graph if

$$(j,k) \notin E \quad \text{if and only if} \quad X_j \perp X_k \mid \{X_i : i \notin \{j,k\}\}.$$

The marginal and conditional independence characterizes different notions of independence. In particular, the marginal independence cannot eliminate the influence of confounding variables and raises questions in interpretation of the results. In contrast, via conditioning on any possible confounding effects, the conditional independence can provide more interpretable results.

To see this, consider random variables $X_1, X_2, X_3$ with an underlying Bayesian network $X_1 \leftarrow X_2 \rightarrow X_3$. Here the notation $\rightarrow$ represents a directed edge and is the common notation used in Pearl (2000). It is straightforward that in this example $X_1, X_3$ are marginally dependent while conditionally independent given $X_2$.

Of note, the conditional independence graph can also have more edges than the marginal independence graph. Consider $X_1, X_2, X_3$ with the Bayesian network $X_1 \rightarrow X_2 \leftarrow X_3$. Then we have $X_1$ and $X_3$ are marginally independent while conditionally dependent given $X_2$.

## 2.2 Impact of Dimension Reduction

Next, we turn to study the impact of dimension reduction on the marginal and conditional independence graphs. To this end, let's consider any dimension reduction procedure mapping $(\widetilde{X}_{C_1}, \ldots, \widetilde{X}_{C_K})^T$ to $\widetilde{Y} = (Y_1, \ldots, Y_K)^T = \{g_1(\widetilde{X}_{C_1}), \ldots, g_K(\widetilde{X}_{C_K})\}^T$. Here $g_j(\cdot) : \mathbb{R}^{d_j} \rightarrow \mathbb{R}$ is a surjective function for each $j \in \{1, \ldots, K\}$.

When $g_1, \ldots, g_K$ are linear functions, we call the corresponding procedure a linear dimension reduction procedure. A linear dimension reduction procedure can be uniquely characterized by a set of mapping directions $\widetilde{w}_1, \ldots, \widetilde{w}_K$ and shift parameters $\widetilde{\beta}_1, \ldots, \widetilde{\beta}_K$. It maps $(\widetilde{X}_{C_1}, \ldots, \widetilde{X}_{C_K})^T$ to $\widetilde{Y} = (Y_1, \ldots, Y_K)^T = \{\widetilde{w}_1^T(\widetilde{X}_{C_1} - \widetilde{\beta}_1), \ldots, \widetilde{w}_K^T(\widetilde{X}_{C_K} - \widetilde{\beta}_K)\}^T$. Such procedures include using the mean, weighted mean, principal components, and latent factors in a factor model.

We first study the marginal independence graph. A model encoding a marginal independence graph is sometimes referred to as a covariance graphical model. Cox and Wermuth (1993), Kauermann (1997), Glonek and McCullagh (1995), Cox (1993), and Drton and Richardson (2008), among many others, studied the properties of models encoding the marginal independence for continuous or discrete data. The next proposition shows that, when we are only interested in studying the marginal independence graph, any aforementioned dimension reduction procedure is at least weakly graph-homotopic. Moreover, there exist dimension reduction procedures that are graph-homotopic.

**Proposition 2.1.** For any set of surjective functions $\{g_1, \ldots, g_K\}$ and any $j \neq k \in \{1, \ldots, K\}$, we have

$$\widetilde{X}_{C_j} \perp \widetilde{X}_{C_k} \quad \text{implies} \quad g_j(\widetilde{X}_{C_j}) \perp g_k(\widetilde{X}_{C_k}).$$

Moreover, if $\widetilde{X}_{C_j} \not\perp \widetilde{X}_{C_k}$, i.e., there exists at least one $i_j \in C_j$ and $i_k \in C_k$ such that $X_{i_j} \not\perp X_{i_k}$, then there exist surjective functions $g_j$ and $g_k$ such that $g_j(\widetilde{X}_{C_j}) \not\perp g_k(\widetilde{X}_{C_k})$.

Proposition 2.1 is a direct consequence of the properties of independence, and the proof is, accordingly, omitted. However, there are two notable remarks. First, there is a natural link between Proposition 2.1 and the researches on modeling the marginal independence. A common approach in modeling the marginal independence is through the latent model. For example, multivariate logistic transformation models are proposed in Glonek and McCullagh (1995) and Kauermann (1997) for modeling the binary data. Considering $\widetilde{X}$ as the latent data and $\widetilde{Y}$ as the observed data, any dimension reduction method corresponds to a latent model. Proposition 2.1 then helps to determine the covariance graphical model and is actually the foundation of many results in modeling marginal independence. See, for example, discussions in Section 5 of Kauermann (1997).

Secondly, it is possible to have $\widetilde{X}_{C_j} \not\perp \widetilde{X}_{C_k}$ but $g_j(\widetilde{X}_{C_j}) \perp g_k(\widetilde{X}_{C_k})$. For example, consider the Gaussian model and linear mappings $g_j(\widetilde{x}) = \widetilde{w}_j^T \widetilde{x}$ ($j = 1, \ldots, K$). Then the above event occurs if and only if $\widetilde{w}_j^T \mathrm{cov}(\widetilde{X}_{C_j}, \widetilde{X}_{C_k}) \widetilde{w}_k = 0$. However, considering that the space of all vectors $(\widetilde{w}_j^T, \widetilde{w}_k^T)^T \in \mathbb{R}^{d_j + d_k}$, those satisfying $\widetilde{w}_j^T \mathrm{cov}(\widetilde{X}_{C_j}, \widetilde{X}_{C_k}) \widetilde{w}_k = 0$ form a strict subspace of $\mathbb{R}^{d_j + d_k}$ when $\widetilde{X}_{C_j} \not\perp \widetilde{X}_{C_k}$. Therefore, if we randomly sample points from $\mathbb{R}^{d_j + d_k}$, it is of probability one that we will have $(\widetilde{w}_j^T, \widetilde{w}_k^T)^T$ such that $\widetilde{w}_j^T \widetilde{X}_{C_j} \perp \widetilde{w}_k^T \widetilde{X}_{C_k}$.

Consider now the settings for the conditional independence graph. That is, the graph characterizing the conditional independence relations between random variables. Such graphs are sometimes referred to as Markov random fields (Lauritzen, 1996). Graph-homotopic mappings are less clear when we consider the conditional independence instead of the marginal independence graph.

On one hand, there are still cases where dimension reduction methods can be graph-homotopic. For example, consider the degenerate graph, i.e., the set of edges $E_C$ is empty. In this case, the conditional independence graph is equivalent to the marginal independence graph. Then Proposition 2.1 applies and any dimension reduction method must also result in an empty set of edges $E_{DR}$.

On the other hand, out of this safe region, there is no theoretical guarantee at all that the dimension reduction procedures are even weakly graph-homotopic. This statement is easy to understand. Consider the conditional independence relations of only three random variables $X_1, X_2, X_3$, and a surjective function $g : \mathbb{R} \to \mathbb{R}$. Then the fact that $X_1 \perp X_2 \mid X_3$ does not necessarily imply $X_1 \perp X_2 \mid g(X_3)$.

In the following, we present a simple example to show that it is possible that the structures of conditional independence graphs, before and after some frequently exploited dimension reduction methods, are very different.

Let's consider the Gaussian graphical model. A Gaussian graphical model defines a multivariate Gaussian distribution with a conditional independence graph. We assume $\widetilde{X} = (\widetilde{X}_{C_1}^T, \ldots, \widetilde{X}_{C_K}^T)^T \sim N_d(0, \Sigma)$ and $\widetilde{X}_{C_i} \in \mathbb{R}^{d_i}$. Here each $\widetilde{X}_{C_i}$ is a cluster with $d_i$ random variables. We are interested in estimating the zero's and nonzero's in the concentration matrix $\Omega = \Sigma^{-1}$ because it encodes the conditional independence graph of $\widetilde{X}$ (Lauritzen, 1996).

For the Gaussian graphical model, it is obvious that a linear dimension reduction procedure is graph-homotopic if and only if the projection directions $\{\widetilde{w}_1, \ldots, \widetilde{w}_K\}$ and shift parameters $\widetilde{\beta}_1, \ldots, \widetilde{\beta}_K$ satisfy the following condition. That is, for the dimension reduced random vector

$$\widetilde{Y} = \{\widetilde{w}_1^T(\widetilde{X}_1 - \widetilde{\beta}_1), \ldots, \widetilde{w}_d^T(\widetilde{X}_d - \widetilde{\beta}_K)\}^T \tag{2.1}$$

with $\Sigma_{\widetilde{Y}} = \mathrm{Cov}(\widetilde{Y})$ and $\Omega_{\widetilde{Y}} = \Sigma_{\widetilde{Y}}^{-1}$, we have $(\Omega_{\widetilde{Y}})_{ij} = 0$ if and only if $\Omega_{C_i, C_j} = 0$. In other words, the dimension reduction procedure needs to preserve the block-wise zero and nonzero patterns of the concentration matrix $\Omega$.

We consider a simple example. The raw random vector $\widetilde{X} = (X_1, X_2, X_3, X_4)^T$ is Gaussian distributed with the concentration matrix $\Omega$. We set the vertex sets $A = \{1\}$, $B = \{4\}$, and $C = \{2, 3\}$. The concentration matrix $\Omega$ is set to be

$$\Omega = \begin{pmatrix} 1 & 0\cdot 5 & 0\cdot 5 & 0 \\ 0\cdot 5 & 1 & 0\cdot 5 & 0\cdot 5 \\ 0\cdot 5 & 0\cdot 5 & 1 & 0\cdot 5 \\ 0 & 0\cdot 5 & 0\cdot 5 & 1 \end{pmatrix}.$$

Because the Gaussian graphical model is encoded in the concentration matrix, the vertex sets $A$ and $B$ are conditionally independent given $C$, which can be described by a conditional independence graph. Accordingly, if the dimension reduction procedure is graph-homotopic, then the corresponding dimension reduced random vector $\widetilde{Y}$ should inherit the graphical structure of $\widetilde{X}$. In comparison, after the dimension reduction using the mean, by simple algebra, we have the dimension reduced random vector $\widetilde{Y}_m = \{X_1, (X_2 + X_3)/2, X_4\}^T$ has the concentration matrix

$$\Omega_{\widetilde{Y}_m} = \begin{pmatrix} 1 & 1 & -1 \\ 1 & 3 & -2 \\ -1 & -2 & 2 \end{pmatrix}.$$

This corresponds to a fully connected graph and the strength of the edge between the first and third vertices is high and not ignorable. The same phenomenon takes place while exploiting principal component analysis. After the dimension reduction using principal component analysis, the dimension reduced random vector $\widetilde{Y}_{PC} = \{X_1, (X_2 + X_3)/\sqrt{2}, X_4\}^T$ has the concentration matrix

$$\Omega_{\widetilde{Y}_{PC}} = \begin{pmatrix} 1\cdot 00 & 0\cdot 71 & -0\cdot 71 \\ 0\cdot 71 & 1\cdot 50 & -0\cdot 79 \\ -0\cdot 71 & -0\cdot 79 & 1\cdot 08 \end{pmatrix}.$$

These show that conditional independence graphs of both $\widetilde{Y}_m$ and $\widetilde{Y}_{PC}$ are distinct from the conditional independence graph of $\widetilde{X}$. This verifies that for certain graphical models the prevailing linear dimension reduction methods such as using the mean or principal component analysis cannot be graph-homotopic.

## 2.3 An Empirical Evaluation of the Impact

In this section, we evaluate the impact of dimension reduction on graphs using the ADHD-200 dataset (Biswal et al., 2010). This dataset provides resting state functional magnetic resonance images of 478 healthy subjects. There are 282,796 voxels in each image. Following Power et al. (2011), we parcellate each image into 264 regions of interest. We are interested in the marginal and conditional independence graphs for the 264 regions. We investigate three dimension reduction methods for regional summarization, namely the median, principal component analysis, and maximum likelihood factor analysis. For estimating the conditional independence graph, we adopt the neighborhood selection method (Meinshausen and Bühlmann, 2006). For estimating the marginal independence graph, we use the covariance thresholding method (Karoui, 2008; Bickel and Levina, 2008). The tuning parameters in neighborhood selection and covariance thresholding are selected using the stability approach (Liu et al., 2010).

Table 1: Comparison of the impact of three dimension reduction methods in estimating the marginal and conditional independence graphs. Each value in the table represents the discrepancy rate of the corresponding two methods. For example, 0·18 in the first row and second column means that in the conditional independence graph estimated using the median for dimension reduction, 18% of the edges are not in the graph estimated using principal component analysis for dimension reduction

|  | conditional independence graph | | | marginal independence graph | | |
| --- | --- | --- | --- | --- | --- | --- |
|  | median | PCA | factor | median | PCA | factor |
| median | 0·00 | 0·18 | 0·19 | 0·00 | 0·07 | 0·07 |
| PCA | 0·15 | 0·00 | 0·17 | 0·05 | 0·00 | 0·05 |
| factor | 0·15 | 0·15 | 0·00 | 0·05 | 0·05 | 0·00 |

For two dimension reduction methods, $M_1$ and $M_2$, we describe the difference in the corresponding graphs by the discrepancy rate $r(M_1, M_2)$. This is defined as

$$r(M_1, M_2) = \frac{\text{CARD}(E_{M_1} \setminus E_{M_2})}{\text{CARD}(E_{M_1})}.$$

Here for a set $E$, $\text{CARD}(E)$ represents the cardinality of $E$. For any dimension reduction method $M$, $E_M$ represents the edge set of the corresponding graph. Table 1 provides the discrepancy rate for each two of the three dimension reduction methods. We observe that the conditional independence graphs corresponding to different dimension reduction methods are very different from each other. This implies that dimension reduction itself can seriously affect the corresponding graphs. In comparison, the marginal independence graphs are much closer to each other. This coincides with the theories. Of note, the three marginal independence graphs are not identical. This indicates potential misidentification of the regions of interest.

## 3 Graph-Homotopic Mappings under Gaussian Graphical Models

In this section we study the existence of graph-homotopic mappings for non-degenerate conditional independence graphs. We focus on the Gaussian graphical models. Compared to the nonlinear methods, linear dimension reduction is more natural for studying the Gaussian data (Fodor, 2002). Therefore, we only consider linear methods in this section.

Fodor (2002) summarized four types of linear dimension reduction methods, namely principal component analysis, factor analysis, independent component analysis, and projection pursuit. Under Gaussian assumptions, the estimands of the four methods can all be interpreted as latent factors in a factor model.

In the following we study the factor model. We characterize the necessary and sufficient conditions such that the mappings from the observed data to latent factors are graph-homotopic or weakly graph-homotopic. Furthermore, we introduce two sub-models of the general factor model, the spectral chord model and the latent rotation model. The former is a factor model with uncorrelated noises, and the later is designed for principal component analysis. The latent rotation model can relax the uncorrelatedness assumption on the noisy terms by using additional properties of the principal components.

In the sequel, for any matrices $M_1, \ldots, M_K$, we denote $\text{diag}(M_1, \ldots, M_K)$ to be

$$\text{diag}(M_1, \ldots, M_K) = \begin{pmatrix} M_1 & 0 & \ldots & 0 \\ 0 & M_2 & \ldots & 0 \\ & & \ddots & \\ 0 & 0 & \ldots & M_K \end{pmatrix}.$$

We first focus on the general factor model corresponding to the data structure in Section 2.1. In particular, suppose $\widetilde{X}$ has $K$ blocks, $\widetilde{X} = (\widetilde{X}_{C_1}, \ldots, \widetilde{X}_{C_K})^T$, where $\widetilde{X}_{C_i} \in \mathbb{R}^{d_i}$, and $\sum_{i=1}^{K} d_i = d$. We assume that each block of $\widetilde{X}_{C_i}$ is generated from a factor model

$$\widetilde{X}_{C_i} = A_i Y_i + \widetilde{\epsilon}_i \quad (i = 1, \ldots, n), \tag{3.1}$$

where $A_i \in \mathbb{R}^{d_i \times 1}$ is the factor loading matrix, $Y_i \in \mathbb{R}$ is the latent factor, and $\widetilde{\epsilon}_i \in \mathbb{R}^{d_i}$ is the noisy term independent of $\{Y_i, i = 1, \ldots, n\}$. The noise term $\widetilde{\epsilon}_i$ is sometimes referred as the idiosyncratic error. Equation (3.1) is equivalent to the following form.

$$\widetilde{X} = A\widetilde{Y} + \widetilde{\epsilon}, \tag{3.2}$$

where $A = \text{diag}(A_1, \ldots, A_K) \in \mathbb{R}^{d \times K}$, $\widetilde{Y} = (Y_1, \ldots, Y_K)^T \in \mathbb{R}^K$, and $\widetilde{\epsilon} = (\widetilde{\epsilon}_1^T, \ldots, \widetilde{\epsilon}_K^T)^T \in \mathbb{R}^d$. We further assume that $\widetilde{X} \sim N_d(0, \Sigma)$, $\widetilde{Y} \sim N_K(0, \Sigma_{\widetilde{Y}})$, and $\widetilde{\epsilon} \sim N_d(0, \Psi)$ where $\Psi = \text{Cov}(\widetilde{\epsilon})$ is the covariance matrix of $\widetilde{\epsilon}$ with diagonals $\Psi_i = \text{Cov}(\widetilde{\epsilon}_i) \in \mathbb{R}^{d_i \times d_i}$ $(i = 1, \ldots, K)$. Here $\widetilde{X}$ can be regarded as the raw data and $\widetilde{Y}$ as the dimension reduced data.

**Remark 3.1.** Principal component analysis, factor analysis, independent component analysis, and projection pursuit can all be regarded as estimating latent factors in a general factor model. Take principal component analysis for example. Let $\widetilde{X}^* \in \mathbb{R}^{d^*}$ be a random vector with covariance matrix $\Sigma^*$, and let $U^* = (\widetilde{u}_1^*, \ldots, \widetilde{u}_{d^*}^*)$ consist of eigenvectors $\widetilde{u}_1^*, \ldots, \widetilde{u}_{d^*}^*$ of $\Sigma^*$ corresponding to the eigenvalues $\lambda_1^* \geq \lambda_2^* \geq \cdots \geq \lambda_{d^*}^*$. The principal components of $\widetilde{X}^*$ are $\widetilde{Y}^* = (Y_1^*, \ldots, Y_{d^*}^*)^T$ with $Y_i^* = \widetilde{u}_i^{*T} \widetilde{X}^*$. We can then write $\widetilde{Y}^* = U^{*T} \widetilde{X}^*$, or equivalently,

$$\widetilde{X}^* = \widetilde{u}_1^* Y_1^* + \sum_{i=2}^{d^*} \widetilde{u}_i^* Y_i^*. \tag{3.3}$$

Consider $Y_1^*$ to be the parameter of interest. Equation (3.3) can be interpreted as a factor model, with the latent factor $Y_1^*$, factor loading matrix $\widetilde{u}_1^*$, and noisy term $\sum_{i=2}^{d^*} \widetilde{u}_i^* Y_i^*$. Under the Gaussian model, it is straightforward to check that Equation (3.3) satisfies the conditions in Equation (3.1). Therefore, the principal components are latent factors of a specific factor model.

In the following, we study the general factor model in Equation (3.2). We characterize necessary and sufficient conditions such that $\widetilde{X}$ and $\widetilde{Y}$ encode consistent conditional independence graphs. In other words, the linear dimension reduction procedure mapping $\widetilde{X} \in \mathbb{R}^d$ to $\widetilde{Y} \in \mathbb{R}^K$ is graph-homotopic.

**Theorem 3.2.** Assume the factor model in Equation (3.2) holds. We then have the linear dimension reduction procedure mapping $\widetilde{X}$ to $\widetilde{Y}$ is graph-homotopic if and only if

$$(\Sigma_{\widetilde{Y}}^{-1})_{ij} = 0 \iff \{(A\Sigma_{\widetilde{Y}}A^T + \Psi)^{-1}\}_{ij} = 0. \tag{3.4}$$

In particular, when $\Psi = \text{diag}(\Psi_1, \ldots, \Psi_K)$ is block diagonal and invertible, the above mapping is graph-homotopic if and only if

$$(\Sigma_{\widetilde{Y}}^{-1})_{ij} = 0 \iff \{(\Sigma_{\widetilde{Y}} + \Phi)^{-1}\}_{ij} = 0, \tag{3.5}$$

where $\Phi = (A^T\Psi^{-1}A)^{-1} = \text{diag}\{1/(A_1^T\Psi_1 A_1), \ldots, 1/(A_K^T\Psi_K A_K)\}$.

Theorem 3.2 characterizes conditions for the latent factors in Equation (3.2) to inherit the conditional independence structure of the raw data. In the following we consider a simplified factor model of Equation (3.2) such that the conditions in Theorem 3.2 hold, so that the latent factors inherit the structure of the raw data.

Let $\Sigma_{\widetilde{Y}}$'s eigen decomposition be

$$\Sigma_{\widetilde{Y}} = U\Lambda_{\widetilde{Y}}U^T, \text{ with } U = (\widetilde{U}_1, \ldots, \widetilde{U}_K), \ \Lambda_{\widetilde{Y}} = \text{diag}(\lambda_1, \ldots, \lambda_K), \text{ and } \lambda_1 \geq \lambda_2 \geq \cdots \geq \lambda_K.$$

We then consider the following spectral chord model. It restricts the eigenvectors of $\Sigma_{\widetilde{Y}}$ in the model depicted in Equation (3.2).

**Definition 3.1.** The factor model in Equation (3.2) follows the spectral chord model if and only if we have

1. The factor loading matrices $A_1, A_2, \ldots, A_K$ are identical;
2. The noisy terms $\widetilde{\epsilon}_1, \widetilde{\epsilon}_2, \ldots, \widetilde{\epsilon}_K$ are independently and identically distributed;
3. If $\{(\Sigma_{\widetilde{Y}} + \Phi)^{-1}\}_{jk} = 0$, then $(\widetilde{U}_i)_j(\widetilde{U}_i)_k = 0$ $(i = 1, \ldots, K)$.

The next theorem shows that, under the spectral chord model, the latent factors $Y_1, \ldots, Y_K$ inherit the conditional independence structure of $\widetilde{X}_{C_1}, \ldots, \widetilde{X}_{C_K}$.

**Theorem 3.3.** Under the spectral chord model, we have if $(\Sigma^{-1})_{jk} = 0$, then $(\Sigma_{\widetilde{Y}}^{-1})_{jk} = 0$. Accordingly, because the concentration matrix of a Gaussian distribution encodes the conditional independence structure, the dimension reduction procedure mapping $\widetilde{X}$ to $\widetilde{Y}$ is weakly graph-homotopic.

Next we investigate the principal component analysis for dimension reduction. We propose an alternative to the spectral chord model, named the latent rotation model, such that principal component analysis gives a graph-homotopic mapping. To this end, we consider the following latent variable model. Suppose we have clean latent random vectors $\widetilde{Z}_1, \ldots, \widetilde{Z}_K$ that are jointly Gaussian distributed, each with dimension $d_f$. Consider $A = (a_{jk}) \in \mathbb{R}^{K \times K}$ to be a symmetric invertible matrix with nonnegative entries, and $\{D_j\}_{j=1}^K \subset \mathbb{R}^{d_f \times d_f}$ to be a collection of diagonal matrices with nonzero diagonal values. The latent random vectors are called clean because we force them to satisfy the following two assumptions.

Assumption 1. For $j \in \{1, \ldots, K\}$, $\text{Cov}(\widetilde{Z}_j) = a_{jj} D_j^2$ is a diagonal matrix.

Assumption 2. For $j, k \in \{1, \ldots, K\}$, $\text{Cov}(\widetilde{Z}_j, \widetilde{Z}_k) = a_{jk} D_j D_k$ is a diagonal matrix.

Consider these two assumptions. The first states that the entries in $\widetilde{Z}_j$ are independent of each other, and the second states that any two latent random vectors, $\widetilde{Z}_j$ and $\widetilde{Z}_k$, are only correlated within the corresponding entries. Here $\widetilde{Z}_j$ could have arbitrary diagonal covariance matrix, while the covariance matrix between $\widetilde{Z}_j$ and $\widetilde{Z}_k$ needs to satisfy the constraints in Assumption 2.

The observed random vector, $\widetilde{X} = (\widetilde{X}_{C_1}^T, \ldots, \widetilde{X}_{C_K}^T)^T \in \mathbb{R}^d$ with $d = d_f K$, is generated via rotating the latent vectors $\widetilde{Z}_1, \ldots, \widetilde{Z}_K$. More specifically, assume that $U_1, \ldots U_K \in \mathbb{R}^{d_f \times d_f}$ are orthogonal matrices. The observed random vector $\widetilde{X}$ is generated via rotating the latent random vectors $\{\widetilde{Z}_j\}_{j=1}^K$ with $\widetilde{X}_{C_j} = U_j \widetilde{Z}_j \in \mathbb{R}^{d_f}$. If $\widetilde{X}$ can be generated in this way, then $\widetilde{X}$ follows the latent rotation model. The following theorem shows that the dimension reduction procedure using principal component analysis is graph-homotopic under the latent rotation model.

**Theorem 3.4.** Suppose that $\widetilde{X} = (\widetilde{X}_{C_1}^T, \ldots, \widetilde{X}_{C_K}^T)^T$ follows the latent rotation model and the projection directions $\widetilde{w}_1, \ldots, \widetilde{w}_K$ in Equation (2.1) are the eigenvectors, not necessarily the leading ones, of $\text{Cov}(\widetilde{X}_{C_1}), \ldots, \text{Cov}(\widetilde{X}_{C_K})$. Then the dimension reduction procedure using $\widetilde{w}_1, \ldots, \widetilde{w}_K$ and arbitrary shift parameters $\widetilde{\beta}_1, \ldots, \widetilde{\beta}_K$ is graph-homotopic.

# 4 Discussion

## 4.1 Main Messages

The latent rotation model and the spectral chord model are arguably difficult to verify. However, they do shed light on the existence of graph-homotopic mappings. In practice, in light of the results in this paper, there are two main suggestions we give to the researchers when the graph of interest is conditional independence graph.

- First, a preliminary analysis on the data for meaningful dimension reduction is necessary. For example, in light of Theorem 3.4, we recommend the researchers to use first and second principal components separately for dimension reduction. If the graph estimate based on first principal components is very different from the graph estimate based on second principal components, then the latent rotation model is very likely to be violated.

- Secondly, if the researchers are confident that the latent rotation model or the spectral chord model does not hold for the data of interest, we recommend a more delicate data analysis. More specifically, instead of separately conducting dimension reduction and graphical model estimation, we recommend to formulate a one step procedure for simultaneously conducting dimension reduction and graphical model estimation. Such a procedure is computationally significantly heavier, but well suited for more general settings.

## 4.2 Relevance to Graph Homotopy in Algebraic Combinatorics

There is a notion of graph homotopy in the literature on algebraic combinatorics. It was defined and exploited in Barcelo et al. (2001) and Babson et al. (2006) for studying the homotopy of graph homomorphisms and homotopy equivalence of graphs (Babson et al., 2006). This notion of homotopy equivalence of graphs should not be confused with the concept of graph-homotopic mapping proposed in this paper.

There are two key differences. (i) The mappings of interest are different. Theirs are those from graphs to graphs, while ours are from Euclidean space to Euclidean space. (ii) The graphs of interest are also different. They study general graphs with no specific relations between vertices, while we study those depicting marginal or conditional independence relations between random variables corresponding to vertices in the graph.

Of separate interest, using the notion of graph-homotopy equivalence introduced in Definition 5.2 in Barcelo et al. (2001), it is straightforward to show that $G_R$, the graph of the raw data, and $G_{DR}$, the graph of the dimension reduced data, are actually graph-homotopy equivalent. However, such graph-homotopy equivalence gives little information towards studying the graph-homotopic mappings.

## 4.3 Impact of Variable Transformation on Graphs

In real applications it is often the case where we do not observe the raw data but only a transformed one. In this section we summarize and focus on four families of transformations. They are marginal and multivariate linear transformations, and marginal and multivariate nonlinear transformations. To be illustrative, we assume the raw data to be Gaussian distributed and we are interested in studying the relations between entries of $\widetilde{X} = (X_1, \ldots, X_d)^T \in \mathbb{R}^d$ with $\Sigma = \text{cov}(\widetilde{X})$. It is readily verified that the impacts of different transformations on the marginal and conditional independence graphs are as follows.

- For marginal linear transformations mapping $\widetilde{X} = (X_1, \ldots, X_d)^T$ to $\widetilde{Y} = (a_1 + b_1 X_1, \ldots, a_d + b_d X_d)^T$ for some $a_i \in \mathbb{R}$ and $b_i \in \mathbb{R} \setminus \{0\}$ ($i = 1, \ldots, d$), neither marginal nor conditional independence graph is affected by such marginal linear transformations.

- For multivariate linear transformations mapping $\widetilde{X}$ to $\widetilde{Y} = \widetilde{\mu} + A\widetilde{X}$ for some $\widetilde{\mu} \in \mathbb{R}^d$ and $A \in \mathbb{R}^{d \times d}$, the necessary and sufficient conditions for $\widetilde{Y}$ to have the same marginal or conditional independence structure as $\widetilde{X}$ is that $A\Sigma A^T$ or $(A\Sigma A^T)^{-1}$ has the same zero-nonzero pattern as $\Sigma$ or $\Sigma^{-1}$. In particular, there exist cases where $\widetilde{Y}$ inherits neither marginal nor conditional independence structure of $\widetilde{X}$.

- For marginal nonlinear transformations mapping $\widetilde{X}$ to $\widetilde{Y}$ with $Y_i = f_i(Y_i)$ ($i = 1, \ldots, d$) for some real nonlinear functions $f_i(\cdot) : \mathbb{R} \to \mathbb{R}$, on one hand, $\widetilde{Y}$ always inherits the marginal independence structure of $\widetilde{X}$. On the other hand, when $\{f_i(\cdot), i = 1, \ldots, d\}$ are bijective functions, $\widetilde{Y}$ inherits the conditional independence structure of $\widetilde{X}$. When $\{f_i(\cdot), i = 1, \ldots, d\}$ are surjective but not bijective, there exist cases such that $\widetilde{Y}$ does not inherit the conditional independence structure of $\widetilde{X}$.

- For multivariate nonlinear transformations $f(\cdot) : \mathbb{R}^d \to \mathbb{R}^d$, there exist cases such that $\widetilde{Y}$ inherits neither marginal nor conditional independence structure of $\widetilde{X}$.

## 4.4 Discussion of Future Studies

In this section, we provide a discussion and list directions for investigation in future studies. First, we conjecture that for each dimension reduction procedure, there should be a nontrivial subset of graphical models under which the procedure is graph-homotopic. Confined in the Gaussian graphical model, this paper characterizes such sets of graphical models for a large family of linear dimension reduction methods. Characterization of such sets of graphical models for nonlinear dimension reduction methods will be studied in the future.

Secondly, in low dimensions, the spectral chord model and latent rotation model discussed in Section 3 are testable. However, in high dimensions, how to develop efficient and theoretically justifiable testing statistics for these models is of strong methodological and theoretical interest. We will leave this for future investigation.

Of note, throughout this paper our main focus is confined in the population level. We do not discuss the empirical implementations. For example, in practice, the leading eigenvector of each cluster $\widetilde{X}_{C_j}$ needs to be estimated. Under mild conditions, Mendelson and Paouris (2012) showed the leading eigenvector of the sample covariance matrix attains the rate of convergence $O_P\{(d/n)^{1/2}\}$ in approximating its population counterpart. We can further accelerate the rate of convergence to $O_P[\{s\log(d/s)/n\}^{1/2}]$, where $s$ represents the number of nonzero entries in the population eigenvector, using sparse principal component analysis (Yuan and Zhang, 2013; Vu et al., 2013) under more stringent assumptions on the sparsity of the eigenvectors. When principal component analysis or sparse principal component analysis provides consistent estimators, the graphical structure can be approximately recovered in the asymptotic sense under the latent rotation model.

Nonetheless, we note the main thrust of this manuscript is to raise the issue of the importance of graph-homotopic dimension reduction in order to allow a contextually accurate interpretation of results, when a researcher is willing to condition on the correctness of the subgraph locations. Analytic issues are only further compounded if this starting point is not stipulated. Thus they are not covered in this manuscript.

We conclude with a discussion on the implications in brain connectivity analysis. In this area, there is substantial uncertainty over node definition and the use of an appropriate collection of parcellations or seeds to perform analyses. This is compounded with the statistical community's promotion of multivariate graphical techniques in the area. Thus it is troubling that, even with a correct and known parcellation scheme with a clear graphical structure adhering to this scheme, this graph can be lost with dimension reduction. This then supports further research into joint analyses uncovering the parcellation schemes and graphs, such as the latent hierarchal Gaussian graphical model proposed in Luo (2014).

## Acknowledgement

We thank Xi Luo and Moo K. Chung for very helpful discussions. We thank the Editor, the Associate Editor, and the referee for very constructive comments.

# A  Technical Proofs

## A.1  Proof of Theorem 3.2

*Proof.* Taking covariance on both sides of Equation (3.2), we have

$$\Sigma = A\Sigma_{\widetilde{Y}}A^T + \Psi.$$

Thus, Equation (3.4) is obvious. We then prove Equation (3.5). When $\Psi$ is block diagonal, by Woodbury formula (Hager, 1989), we have

$$\Sigma^{-1} = (A\Sigma_{\widetilde{Y}}A^T + \Psi)^{-1} = \Psi^{-1} - \Psi^{-1}A(\Sigma_{\widetilde{Y}}^{-1} + \Phi^{-1})^{-1}A^T\Psi^{-1}.$$

Since $\Psi^{-1}$ and $A$ are block diagonal, we have $(\Sigma^{-1})_{C_i,C_j} = 0$ if and only if $(\Sigma_{\widetilde{Y}}^{-1} + \Phi^{-1})^{-1}_{ij} = 0$ ($i,j = 1,\ldots,K$). Again using Woodbury formula, we have

$$(\Sigma_{\widetilde{Y}}^{-1} + \Phi^{-1})^{-1} = \Phi - \Phi(\Sigma_{\widetilde{Y}} + \Phi)^{-1}\Phi.$$

The desired result follows by noting that $\Phi$ is diagonal. $\square$

## A.2  Proof of Theorem 3.3

*Proof.* By the property of the spectral chord model, we have

$$A_1 = A_2 = \cdots = A_K, \quad \Psi_1 = \Psi_2 = \cdots = \Psi_K.$$

Accordingly, in Equation (3.5), we have

$$\Phi = \operatorname{diag}\{1/(A_1^T\Psi_1 A_1),\ldots,1/(A_K^T\Psi_K A_K)\} = \sigma^2 I_d, \quad \sigma^2 = 1/(A_1^T\Psi_1 A_1).$$

This implies that, under the spectral chord condition for the eigenvectors of $\Sigma_{\widetilde{Y}}$, for those satisfying $\{(\Sigma_{\widetilde{Y}} + \Psi)^{-1}\}_{jk} = 0$, we have

$$(\Sigma_{\widetilde{Y}}^{-1})_{jk} = \sum_{i=1}^K \frac{1}{\lambda_i}(\widetilde{U}_i\widetilde{U}_i^T)_{jk} = \sum_{i=1}^K \frac{1}{\lambda_i}(\widetilde{U}_i)_j(\widetilde{U}_i)_k = 0.$$

It further yields that $E_{DR} \subset E_C$ and the dimension reduction procedure mapping $\widetilde{X}$ to $\widetilde{Y}$ is weakly graph-homotopic. $\square$

## A.3  Proof of Theorem 3.4

Theorem 3.4 is a consequence of the following theorem, which characterizes general sufficient conditions under which the dimension reduction procedure using principal component analysis is graph-homotopic. In the sequel, for any matrix $M \in \mathbb{R}^{d\times d}$ and $i,j \in \{1,\ldots,d\}$, we denote $M_{*j}$ and $M_{i*}$ to be the $j$th column and $i$th row of $M$. It is obvious that the shift parameters $\widetilde{\beta}_1,\ldots,\widetilde{\beta}_K$ do not affect the graphs. Therefore in the sequel, without loss of generality, we assume $\widetilde{\beta}_1 = \cdots = \widetilde{\beta}_K = 0$.

**Theorem A.1.** For the Gaussian random vector $\widetilde{X} \sim N_d(0, \Sigma)$, assume the concentration matrix $\Omega$ has the following structure

$$\Omega = \begin{pmatrix} U_1 \Lambda_{11} U_1^T & U_1 \Lambda_{12} U_2^T & \ldots & U_1 \Lambda_{1K} U_K^T \\ U_2 \Lambda_{21} U_1^T & U_2 \Lambda_{22} U_2^T & \ldots & U_2 \Lambda_{2K} U_K^T \\ \cdot & \cdot & \ddots & \cdot \\ U_K \Lambda_{K1} U_1^T & U_K \Lambda_{K2} U_2^T & \ldots & U_K \Lambda_{KK} U_K^T \end{pmatrix}$$

$$= \begin{pmatrix} U_1 & 0 & \ldots & 0 \\ 0 & U_2 & \ldots & 0 \\ \cdot & \cdot & \ddots & \cdot \\ 0 & 0 & \ldots & U_K \end{pmatrix} \underbrace{\begin{pmatrix} \Lambda_{11} & \Lambda_{12} & \ldots & \Lambda_{1K} \\ \Lambda_{21} & \Lambda_{22} & \ldots & \Lambda_{2K} \\ \cdot & \cdot & \ddots & \cdot \\ \Lambda_{K1} & \Lambda_{K2} & \ldots & \Lambda_{KK} \end{pmatrix}}_{\Lambda} \begin{pmatrix} U_1^T & 0 & \ldots & 0 \\ 0 & U_2^T & \ldots & 0 \\ \cdot & \cdot & \ddots & \cdot \\ 0 & 0 & \ldots & U_K^T \end{pmatrix}, \quad (A.1)$$

where $U_1, \ldots, U_K$ corresponding to the $K$ clusters are invertible, and for any $j, k \in \{1, \ldots, K\}$, $\Lambda_{jk} \in \mathbb{R}^{d_j \times d_k}$ satisfies that

$$(\Lambda_{jk})_{1,\{2,\ldots,d_k\}} = 0, \quad (\Lambda_{jk})_{\{2,\ldots,d_j\},1} = 0. \tag{A.2}$$

In this setting, we have that the linear dimension reduction method with the projection directions $\{\widetilde{w}_1, \ldots, \widetilde{w}_K\}$ satisfying that

$$\widetilde{w}_j = (U_j)_{*1} \quad (j = 1, \ldots, K),$$

is weakly graph-homotopic. Moreover, under the condition that $(\Lambda_{ij})_{11} = 0$ only if $\Lambda_{ij} = 0$, it is a graph-homotopic mapping.

*Proof.* Remind that $\widetilde{X} = (X_1, \ldots, X_d)^T = (\widetilde{X}_{C_1}^T, \ldots, \widetilde{X}_{C_K}^T)^T \in \mathbb{R}^d$. For notational simplicity, we write $\widetilde{X}_j = \widetilde{X}_{C_j}$ ($j = 1, \ldots, K$). Then we have that the dimension reduced random vector $\widetilde{Y}$ can be written as

$$\widetilde{Y} = (Y_1, \ldots, Y_K)^T = \begin{pmatrix} \widetilde{w}_1^T & 0 & \ldots & 0 \\ 0 & \widetilde{w}_2^T & \ldots & 0 \\ \cdot & \cdot & \ddots & \cdot \\ 0 & 0 & \ldots & \widetilde{w}_K^T \end{pmatrix} \begin{pmatrix} \widetilde{X}_1 \\ \widetilde{X}_2 \\ \vdots \\ \widetilde{X}_K \end{pmatrix} \in \mathbb{R}^K.$$

We augment $\widetilde{Y}$ to be $\widetilde{Y}^{\text{AUG}} \in \mathbb{R}^d$ such that

$$\widetilde{Y}^{\text{AUG}} = \begin{pmatrix} U_1^T & 0 & \ldots & 0 \\ 0 & U_2^T & \ldots & 0 \\ \cdot & \cdot & \ddots & \cdot \\ 0 & 0 & \ldots & U_K^T \end{pmatrix} \begin{pmatrix} \widetilde{X}_1 \\ \widetilde{X}_2 \\ \vdots \\ \widetilde{X}_K \end{pmatrix} \in \mathbb{R}^d,$$

where we remind that $U_j = (\widetilde{w}_j, *)$. Here and after $*$ represents the part that we do not specify in the equation. Hence $(\widetilde{Y}_{C_j}^{\text{AUG}})_1 = Y_j$ ($j = 1, \ldots, d$). Accordingly, noting that the density function of $\widetilde{X}$ can be written as

$$f_{\widetilde{X}}(\widetilde{x}) \propto \exp\Big(-\frac{1}{2} \sum_{i,j=1}^{K} \widetilde{x}_i^T \Omega_{C_i, C_j} \widetilde{x}_j\Big),$$

where $\widetilde{x} = (\widetilde{x}_1^T, \ldots, \widetilde{x}_K^T)^T$ with $\widetilde{x}_j = \widetilde{x}_{C_j}$ $(j = 1, \ldots, K)$, we have

$$f_{\widetilde{Y}^{\text{AUG}}}(\widetilde{y}^{\text{AUG}}) \propto \exp\left\{-\frac{1}{2}\sum_{i,j=1}^{K} \widetilde{y}_i^{\text{AUG}T} U_i^{-1} \Omega_{C_i,C_j} (U_j^T)^{-1} \widetilde{y}_j^{\text{AUG}}\right\}.$$

Here $\widetilde{y}^{\text{AUG}} = (\widetilde{y}_1^{\text{AUG}T}, \ldots, \widetilde{y}_K^{\text{AUG}T})^T$ with $\widetilde{y}_j^{\text{AUG}} = \widetilde{y}_{C_j}^{\text{AUG}}$ $(j = 1, \ldots, K)$. Due to the structure of $\Omega$ depicted in Equations (A.1) and (A.2), the right hand side of the above equation can be factorized into a term only involving $\widetilde{y} = (y_1, \ldots, y_K)^T$ and another term only involving $\widetilde{y}^{\text{AUG}} \setminus \widetilde{y}$, namely the random variables in $\widetilde{y}^{\text{AUG}}$ but not in $\widetilde{y}$. Thus, by integrating $\widetilde{y}^{\text{AUG}} \setminus \widetilde{y}$ out, we have

$$f_{\widetilde{Y}}(\widetilde{y}) \propto \exp\left[-\frac{1}{2}\sum_{i,j=1}^{K} \{U_i^{-1}\Omega_{C_i,C_j}(U_j^T)^{-1}\}_{11} y_i y_j\right] = \exp\left\{-\frac{1}{2}\sum_{i,j=1}^{K}(\Lambda_{ij})_{11} y_i y_j\right\}.$$

Accordingly, using the fact that $\widetilde{Y}$ is also Gaussian distributed, we have $Y_i$ is independent of $Y_j$ conditioning on others if $\Omega_{C_i,C_j} = 0$. Moreover, if $(\Lambda_{ij})_{11} = 0$ only if $\Lambda_{ij} = 0$, we have $Y_i$ is independent of $Y_j$ conditioning on others only if $\Omega_{C_i,C_j} = 0$. □

As a direct consequence of Theorem A.1, the following corollary provides a sufficient condition such that the dimension reduction method using the first principal component is a graph-homotopic mapping.

**Corollary A.1.** Suppose that the structure depicted in Equations (A.1) and (A.2) in Theorem A.1 holds. If we further have $U_1, \ldots, U_K$ are orthogonal matrices and $\Lambda$ satisfies that $(\Lambda^{-1})_{C_j,C_j}$ is a diagonal matrix with a descending sequence of positive diagonal values for any $j \in \{1, \ldots, K\}$, then the linear dimension reduction method using the first principal components on each cluster $\widetilde{X}_{C_j}$ is weakly graph-homotopic. Moreover, under the condition that $(\Lambda_{ij})_{11} = 0$ if and only if $\Lambda_{ij} = 0$, it is a graph-homotopic mapping.

*Proof.* Using Equation Equation (A.1), and the fact that $U_1, \ldots, U_K$ are orthogonal matrices, we have

$$\Sigma = \Omega^{-1} = \begin{pmatrix} U_1(\Lambda^{-1})_{C_1,C_1}U_1^T & & & * \\ & U_2(\Lambda^{-1})_{C_2,C_2}U_2^T & & \\ & & \ddots & \\ * & & & U_K(\Lambda^{-1})_{C_K,C_K}U_K^T \end{pmatrix}.$$

Since $(\Lambda^{-1})_{C_j,C_j}$ is a diagonal matrix with descending diagonal values, $U_j$ contains the eigenvectors of $\Sigma_{C_j,C_j}$ and $\widetilde{w}_j$ is its leading eigenvector. Using Theorem A.1, we have the desired result. □

Now we are ready to prove Theorem 3.4.

*of Theorem 3.4.* Under the latent rotation model, the covariance matrix of $(\widetilde{Z}_1^T, \ldots, \widetilde{Z}_K^T)^T$ is

$$\text{Cov}\{(\widetilde{Z}_1^T, \ldots, \widetilde{Z}_K^T)^T\} = \begin{pmatrix} D_1 & 0 & \ldots & 0 \\ 0 & D_2 & \ldots & 0 \\ . & . & \ddots & . \\ 0 & 0 & \ldots & D_K \end{pmatrix} (A \otimes I_{d_f}) \begin{pmatrix} D_1 & 0 & \ldots & 0 \\ 0 & D_2 & \ldots & 0 \\ . & . & \ddots & . \\ 0 & 0 & \ldots & D_K \end{pmatrix},$$

where $\otimes$ represents the tensor product. Accordingly, the covariance matrix $\Sigma$ of $\widetilde{X}$ can be written as

$$\Sigma = \begin{pmatrix} U_1 & 0 & \ldots & 0 \\ 0 & U_2 & \ldots & 0 \\ . & . & \ddots & . \\ 0 & 0 & \ldots & U_K \end{pmatrix} \text{Cov}\{(\widetilde{Z}_1^T, \ldots, \widetilde{Z}_K^T)^T\} \begin{pmatrix} U_1^T & 0 & \ldots & 0 \\ 0 & U_2^T & \ldots & 0 \\ . & . & \ddots & . \\ 0 & 0 & \ldots & U_K^T \end{pmatrix}.$$

Using the fact that for any square invertible matrix $A, B$, we have

$$(A \otimes B)^{-1} = A^{-1} \otimes B^{-1},$$

it is easy to observe that $\Omega = \Sigma^{-1}$ satisfies Equations (A.1) and (A.2). Since $\text{Cov}(\widetilde{Z}_j, \widetilde{Z}_k) = a_{jk} D_j D_k$, $a_{jk} \geq 0$, and $D_j, D_k$ are diagonal matrices with nonzero diagonals, we conclude that $\{\text{Cov}(\widetilde{Z}_j, \widetilde{Z}_k)\}_{11} = 0$ if and only if $\text{Cov}(\widetilde{Z}_j, \widetilde{Z}_k)$ itself is zero. Moreover, because $\text{Cov}(\widetilde{Z}_j, \widetilde{Z}_k)$ is a diagonal matrix, we do not need to specify which eigenvector we exploit as in Corollary A.1. Accordingly, the linear dimension reduction method using any principal component is graph-homotopic. $\square$